\documentstyle[aps,epsfig]{revtex}
\begin{document}
\draft
\title{Isotope effects and the charge gap formation in 
the charge ordered phase of colossal magnetoresistance manganites}
\author{Unjong Yu, Yu. V. Skrypnyk, and B. I. Min}
\address{Department of Physics,
        Pohang University of Science and Technology, 
        Pohang 790-784, Korea}
\date{\today}
\maketitle

\begin{abstract}
Giant oxygen isotope effects observed in colossal    
magnetoresistance manganites are investigated by employing the 
combined model of the double exchange and interacting lattice
polaron mechanism.  We have shown that the isotope effects 
on $ T_C$ in the metallic phase and $ T_{CO}$ in the 
charge ordered phase of manganites can be explained well
in terms of the double exchange and polaron narrowing factors
with reasonable physical parameters.
\end{abstract}

A recent discovery of colossal magnetoresistance (CMR)
phenomena has stimulated enormous interest in the perovskite 
manganites, R$_{1-x}$A$_x$MnO$_3$ 
(RAMO: R = rare-earth; A = divalent cation) \cite{Helmolt,Jin}.
They exhibit a very rich phase diagram depending on  the
doping concentration, temperature, and pressure: antiferromagnetic (AFM) 
insulator, ferromagnetic (FM) metal, charge ordered (CO)
insulator \cite{Schiffer}.
These novel features suggest that several
interactions originating from the spin, charge, and lattice degrees
of freedom are competing. 
For instance, the correlation between ferromagnetism and metallic
conductivity for 0.2 $<$ x $<$ 0.5 was explained qualitatively
in terms of the double exchange (DE) mechanism \cite{Zener}.
On the other hand,  Jahn-Teller polaron effects due to a strong 
electron-phonon interaction are thought to be responsible 
for anomalous properties of manganites\cite{Millis}.

The most evident signature for the lattice degree of freedom is 
considered to be the isotope effect, because it can affect the electronic
properties only through the electron-phonon interaction.
Indeed many recent experiments, such as magnetization,
resistivity, thermal expansion, electron paramagnetic resonance,
and magnetostriction measurements,  have revealed giant isotope effects
both in the metallic and CO phases of RAMO
\cite{Zhao,Sheng,Zhao2,Franck,Zhao3,Isaac,Ibarra,Zhao4,Zhao5,Babu}.
The isotope effect on the magnetic phase transition
temperature $ T_C$ in the metallic phase \cite{Zhao,Sheng,Zhao2,Franck}
and on the charge ordering transition temperature $ T_{CO}$ 
\cite{Zhao3,Isaac,Ibarra,Zhao4,Zhao5,Babu} were confirmed
by replacing $^{16}$O with $^{18}$O.
Zhao {\it et al.} \cite{Zhao} have obtained a shift of 20K in $ T_C$ 
in La$_{0.8}$Ca$_{0.2}$MnO$_3$, yielding the isotope exponent 
$\alpha_c$=0.85 ($\alpha_c \equiv -\Delta\ln T_C/ \Delta\ln M_O$, 
$M_O$: oxygen isotope mass).
On the other hand, Franck {\it et al.} \cite{Franck}
obtained $\alpha_c \sim 0.4$ for the same Ca concentration.
This large difference in $\alpha_c$ between two groups is suspected to arise 
from the different sample stoichiometry due to oxygen excess 
\cite{Franck}.

For the CO phases, Isaac {\it et al.} \cite{Isaac} observed  
$\Delta  T_{CO}$ with a negative oxygen isotope exponent
$\alpha_{co} = -0.41$ in La$_{0.57}$Ca$_{0.43}$MnO$_3$.
Further, novel crossovers from a metallic to a CO insulating ground state
are observed in (La$_{0.5}$Nd$_{0.5}$)$_{2/3}$Ca$_{1/3}$MnO$_3$ 
\cite{Zhao3,Ibarra} and in (La$_{0.175}$Pr$_{0.525}$)Ca$_{0.3}$MnO$_3$ 
\cite{Babu} by the oxygen isotope exchange.
The above systems are metallic with $^{16}$O, but 
are very close to the phase boundary between the FM metal and the CO insulator.
A more interesting feature is manifested in the CO phases 
of Nd$_{0.5}$Sr$_{0.5}$MnO$_3$ and La$_{0.5}$Ca$_{0.5}$MnO$_3$
\cite{Zhao4,Zhao5}, 
which show the strong magnetic field dependence of the isotope effect.
Under the magnetic field of 5.4 T, 
$\Delta  T_{CO}$ increases substantially from 21K to 43K
for Nd$_{0.5}$Sr$_{0.5}$MnO$_3$, and from 9K to 40K for 
La$_{0.5}$Ca$_{0.5}$MnO$_3$, respectively.
Thus, the magnitude of $\alpha_{co} (\equiv -\Delta\ln T_{CO}/ 
\Delta\ln M_O$) increases rapidly 
with the magnetic field. This feature
is in contrast to the case of $\alpha_{c}$ in the metallic phase that
is rather insensitive to the magnetic field \cite{Franck,Zhao5}.
It has been pointed out that these anomalous features are difficult 
to understand within the existing theories of the charge ordering 
transition \cite{Zhao4,Zhao5}.

The isotope effects in the metallic phase can be explained properly 
by using the small polaron model \cite{Zhao}.
In contrast, the understanding of the isotope effects in the CO phase
is not straightforward.
In the previous study \cite{Jdlee2}, we have described successfully
the lattice and magnetic properties of the CO phase of half-doped
Mn-oxides using the combined model of the double exchange and interacting
lattice polaron mechanism. The electron lattice of the CO state
was viewed as the generalized Wigner crystal, that is,
the CO state becomes stable when the
repulsive Coulomb interaction between carriers dominates over
the kinetic energy of carriers. 
In this paper, we have explored the isotope effects in the CO phase of 
CMR manganites on the basis of the similar model.
We have found that the isotope effects and related features
in the CO phase of manganites are well described
in terms of the present model.

Let us first examine the isotope effect in the metallic phase.
The standard small polaron theory shows that the strong electron-phonon
interaction reduces the electron hopping parameter 
by the polaron narrowing factor \cite{Mahan},
\begin{eqnarray}
\tilde{t} = t \; \exp\left[-\sum_{\bf q} |u_{\bf q}|^2 \; (N_q+1/2) \right]
\label{t_tilde}
\end{eqnarray}
with $u_{\bf q} = M_q/\omega_q \left(\exp(i{\bf q}\cdot\bbox{\delta}) 
- 1\right)$.
Here $M_q$ is the parameter of the electron-phonon interaction 
strength, $\omega_q$ is the phonon frequency, 
$N_q$ is the boson distribution function,
and $\bbox{\delta}$ is the displacement to the neighboring site. 
The electron-phonon interaction parameter $M_q$ depends both 
on the ion mass and the phonon frequency 
in the form of $M_q \sim (M_{ ion} \omega)^{-1/2}$.
Hence $\sum |u_{\bf q}|^2$ is proportional to $\sqrt{M_{ ion}}$ 
with the assumption of $\omega \sim (M_{ ion})^{-1/2}$. 
In the DE model,  $ T_C$ is proportional 
to the hopping parameter \cite{Kubo}, {\it i.e.},
$ T_C$ $\sim \tilde{t} \sim t\exp\left[-\gamma\sqrt{M_{ ion}}\right]$,
and so the material with heavier ion mass would have the lower $ T_C$. 
This is nothing but the isotope effect observed in the metallic phase
of RAMO.  Note that, with increasing $\sum_{\bf q} |u_{\bf q}|^2$, 
one gets the larger isotope effect.

The experimental results of Franck {\it et al.} ($\alpha_c = 0.34$) 
\cite{Franck} can be reproduced with $\sum_{\bf q} |u_{\bf q}|^2 = 1.2$ 
and $\omega = 0.07$eV.
In the same framework, one can also explain the reduction of $\alpha_c$ 
with increasing the tolerance factor \cite{Zhao,Franck}. 
The increase of the tolerance factor leads to the 
increase of the Mn$-$O$-$Mn bond angle and the decrease of 
the bond length between neighboring Mn sites \cite{Radaelli}. 
One can thus expect that the phonon frequency becomes hardened 
due to reduced bond length \cite{Ujyu}, and 
accordingly $\sum_{\bf q} |u_{\bf q}|^2$ decreases
considering $\sum_{\bf q} |u_{\bf q}|^2 \sim (\omega)^{-3}$. 
Therefore, we have smaller $\alpha_c$ with increasing the tolerance factor
in agreement with the observation.
In addition, the increase of the Mn$-$O$-$Mn bond angle and 
the decrease of bond length will certainly increase the hopping 
parameter and so weakens the polaron formation \cite{Fehske}.
This effect plays a role of reducing $\sum_{\bf q} |u_{\bf q}|^2$ 
and $\alpha_c$ further.

As for the giant isotope effects in manganites,
Nagaev \cite{Naga} has proposed mechanisms which are not related to the 
electron-phonon interaction. He claimed that a change in the oxygen isotope
leads inevitably to a change in the sample stoichiometry. So thermodynamic
equilibrium densities of oxygen vacancies or excess oxygen
atoms depend on the mass of the oxygen atom and so 
corresponding carrier density has isotope dependence. 
In addition, the oxygen nonstoichiometry (vacancies) produces the isotope
dependent pressure by causing a change in the volume.
These mechanisms seem to explain observed isotope effects
in La$_{0.8}$Ca$_{0.2}$MnO$_3$ which are very sample dependent 
($\alpha_c$=0.85 vs. $\alpha_c \sim 0.4$), as mentioned above.
This scenario, however, works only for the low carrier density regime 
(x $< 3/8$).  Moreover, more recent experiments have shown that 
there are intrinsic isotope effects which are not caused by any possible 
differences in the oxygen content \cite{Franck,Zhao3}. 
Also it is hard to explain with these mechanisms
the isotope effects in the CO phase of manganites.

Now let us consider the isotope effect in the CO phase.
Since $T_{CO}$ is not linearly proportional to the hopping parameter
in this case, 
one has to find the relation between $ T_{CO}$ and the hopping parameter.
The extended polaron Hamiltonian for
the half-doped manganites
incorporating the electron-phonon interaction is expressed as
\begin{eqnarray}
H = \sum_{ij} t^D_{ij} c_{i}^{\dagger} c_j + \frac{1}{2} 
	V_0 \sum_{<i j>} n_i n_j
    + \sum_{\bf q} \omega_q a_{\bf q}^{\dagger} a_{\bf q} 
    + \sum_{j {\bf q}} c_j^{\dagger} c_j e^{i{\bf q} \cdot {\bf R}_j} 
       M_q (a_{\bf q} + a_{-{\bf q}}^{\dagger}),
\end{eqnarray}
where $t^D_{ij} (\equiv t\langle \cos\frac{\theta}{2}\rangle$,
$\theta$: angle between two spins on neighboring sites) is the 
DE hopping parameter and $V_0$ is the Coulomb repulsion 
between two electrons of neighboring sites.  
Using the polaron canonical transformation, one can get
\begin{eqnarray}
\bar{H} = \sum_{ij} \tilde{t}^D_{ij} c_{i}^{\dagger} c_j 
           + \frac{1}{2} V \sum_{<i j>} n_i n_j
           + \sum_{\bf q} \omega_q a_{\bf q}^{\dagger} a_{\bf q}
           - \sum_j n_j \Delta_p,
\end{eqnarray}
where 
$V \left(\equiv V_0 - 2 \sum_{\bf q} \frac{M_q^2}{\omega_q} 
                   \exp(i {\bf q} \cdot \bbox{\delta})\right)$ 
corresponds to the renormalized Coulomb repulsion 
and $\Delta_p (\equiv  \sum_{\bf q} \frac{M_q^2}{\omega_q})$ is the
polaron binding energy. 
The effective hopping $\tilde{t}^D$ is defined as in Eq. (\ref{t_tilde}) 
with the bare hopping $t$ replaced by $t^D$.
It can be shown that $V$ does not depend on the ion mass 
when $\omega \sim (M_{ ion})^{-1/2}$.
Thus the isotope effects will be realized only through 
$\tilde{t}^D$, if they exist.

To describe the CO state in the mean field approximation, 
the size of the unit cell is doubled by introducing formally two identical
sublattices. This procedure automatically implies that the
CO state can be only of a checkerboard type
and certain restrictions are put on the lattice symmetry.
Also, for simplicity, the electron hopping will be permitted only
between sites belonging to different sublattices.
Then the electronic part of the
above Hamiltonian is reduced to the following form:
\begin{eqnarray}
\label{ham}
\bar{H}=\sum_{\alpha\neq \beta}\tilde{t}^D_{\alpha\beta}a_{\alpha}^{\dagger}
a_{\beta}+zV\tau \sum_{\alpha}(-1)^{\alpha}a_{\alpha}^{\dagger }
a_{\alpha},
\end{eqnarray}
where $\alpha$ and $\beta$ are sublattice indices, $z$ is the number 
of nearest neighbors, and $\tau$ is the charge ordering order parameter
($-0.5\le\tau\le 0.5$), describing the deviation in electron 
concentration in the CO phase, {\it i.e.},
$n_{\alpha}=\frac 12+\tau$; $n_{\beta}=\frac 12-\tau$ . 

Diagonalizing the above Hamiltonian,
one gets two states with energy $\pm\epsilon(k)$,
where $\epsilon(k) = \sqrt{ |\tilde{t}^D(k)|^2 + (zV\tau)^2 }$  \cite{Skr}.
Therefore, in the CO phase, the initial band is split into
two bands with a charge gap of $2\Delta=2zV\tau$.
The order parameter $\tau$ can be determined from
the following  BCS-like self-consistent equation,
\begin{eqnarray}
1 = \frac{zV}{2}{1 \over N} 
     \sum_k \frac{\tanh(\beta\epsilon(k)/2)}{\epsilon(k)} ,  \label{sc}
\end{eqnarray}
where $\beta=\frac{1}{k_B T}$.
The wave vector summation in Eq.\ (\ref{sc}) can be replaced by the energy
integration with the appropriate density of states (DOS).
In the case of three-dimensional (3D) system, the DOS without the
charge ordering $\rho_0(\epsilon)$ can be chosen as a semielliptic form, 
$
\rho _0(\varepsilon )=\frac 4{\pi B^2}\sqrt{B^2-\varepsilon ^2},
$
where $B (\equiv z\tilde{t}^D)$ is a half of the bandwidth  
without the charge ordering.
Then the self-consistent equation at $T=0$ reads
$
1=\frac 2\pi \frac{\sqrt{b^2+\tau ^2}}{b^2}\left( K\left( \sqrt{
\frac{b^2}{b ^2+\tau ^2}}\right) -E\left( \sqrt{\frac{b ^2}{
b ^2+\tau ^2}}\right) \right) ,
$
where $b=\frac B{zV}$, and  $K(k)$ and $E(k)$ are 
elliptic integral of the first and the second kind, respectively. 

\begin{figure}
\centerline{\epsfig{figure=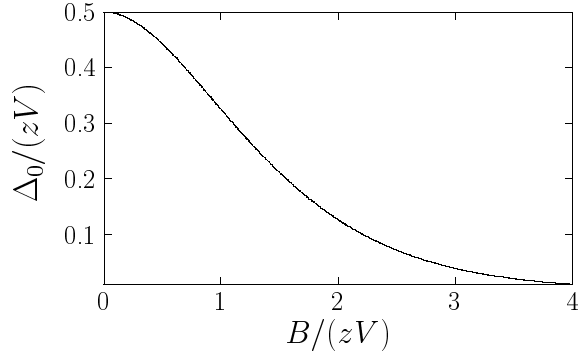,width=8cm}}
\caption{
The zero temperature charge gap $\Delta_0$ as a function of 
the half-bandwidth $B~(=z\tilde{t}^D$)
for the model 3D system. The values are
scaled by $zV$ where $z$ is the number of nearest neighbors and $V$ is the
renormalized Coulomb interaction between carriers. 
}
\label{fig1}
\end{figure}

By expanding the elliptic integrals into series
in both the narrow and wide band width limits,
the analytic form of the charge gap can be obtained:
$\Delta _0\approx 
\frac{zV}2\left( 1-\frac 12\left( \frac B{zV}\right) ^2\right)$ for
$\frac B{zV}\ll 1$, and $ \Delta _0=4Be^{-\left( \frac{\pi B}{2zV}+1\right)}$
for $\frac B{zV}\gg 1.$
Figure~\ref{fig1} provides the overall behavior of $\Delta_0$ as 
a function of the band width. It is seen that $\Delta_0$ diminishes
with increasing the hopping parameter, suggesting
that the transition between the CO and the charge 
disordered state can be described by the competition between the
hopping and the intersite Coulomb repulsion parameter. 

The self-consistent equation Eq. (\ref{sc})  also 
determines the charge ordering temperature $ T_{CO}$.
At $ T_{CO}$, the charge gap is closed in the
spectrum and the order parameter $\tau $ is equal to zero.  
In the narrow band limit,  $\frac B{zV} \ll 1$, the $\tanh$ function
in Eq. (\ref{sc})  can be expanded into the
series. Then $ T_{CO}$ is given by 
$k_B{ T_{CO}} \approx \frac{zV}4\left( 1-\frac 13
\left( \frac B{zV}\right) ^2\right) .
$
Hence for the dispersionless band, it follows that
$\ k_B{ T_{CO}}\approx \frac{zV}4 = \frac{\Delta _0}2$,
in accordance with the previous work \cite{Jdlee2},
In the opposite case of the wide band limit, $\frac B{zV} \gg 1$, 
$ T_{CO}$ can be obtained through the procedure
analogous to the BCS theory \cite{Fett}. 
Careful retaining of all the terms that do not vanish 
under $\frac B{k_B T_{CO}}\rightarrow \infty $ shows that
the BCS result is obtained despite the nonconstant DOS of the 
present band,
$k_B{ T_{CO}}=\frac{e^\gamma }\pi \Delta _0 \approx 0.567 \Delta _0 ,
$
where $\gamma $ is the Euler constant $(\sim 0.5772)$. Thus, the ratio of
$\frac {\Delta _0}{k_B T_{CO}}$  does not change significantly 
with varying the ratio of $\frac B{zV}$.

\begin{figure}
\centerline{\epsfig{figure=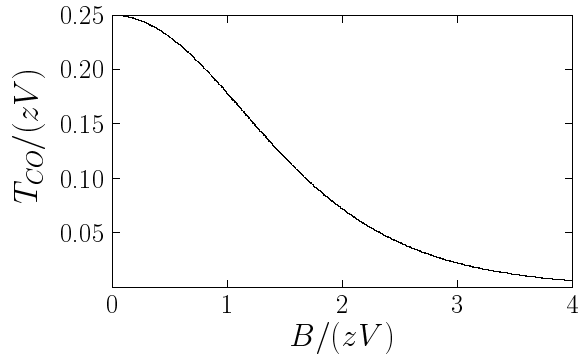,width=8cm}}
\caption{The charge ordering transition temperature
  $T_{CO}$ as a function of half-bandwidth $B$. \label{Tco_B}}
\end{figure}

Overall dependence of $ T_{CO}$
on $\frac B{zV}$ is presented in Fig.~\ref{Tco_B}.
Note that $ T_{CO}$ is $zV/4$ at $B=0$ 
and diminishes as $B$ becomes larger.
The observed decrease of $ T_{CO}$ and the 
concomitant collapse of the CO state in half-doped manganites 
by applying the magnetic field or the
pressure can be qualitatively understood in this context \cite{Tomioka}.
That is, the magnetic field or the pressure increases the bandwidth 
to induce the instability of the CO state.  

Now, based on the above relation of $ T_{CO}$ with the effective hopping 
parameter $\tilde{t}^D$, 
one can account for the isotope effects observed in the CO phase.
Since  $\tilde{t}^D$ has ion mass dependence
in the expression of the polaron narrowing factor, the replacement of
$^{16}$O with heavier $^{18}$O gives rise to a smaller $\tilde{t}^D$
and so yields enhanced $ T_{CO}$. 
Let us estimate the reasonable physical parameters producing
the experimental isotope exponent $\alpha_{co}$.
For a given $ T_{CO} = 150$K, one can have $t^D=0.05$eV, 
$\sum_{\bf q} |u_{\bf q}|^2 = 1.2$, and the screened
Coulomb repulsion $V = 0.018$eV.  The bare hopping parameter $t^D$
in the CO phase is thought to be smaller than that in the metallic 
phase because of the different spin ordering fluctuations. 
With these parameters, 
one could get the isotope exponent $\alpha_{co} = -0.44$
that is consistent with experiments \cite{Isaac,Zhao5}.

\begin{figure}
\centerline{\epsfig{figure=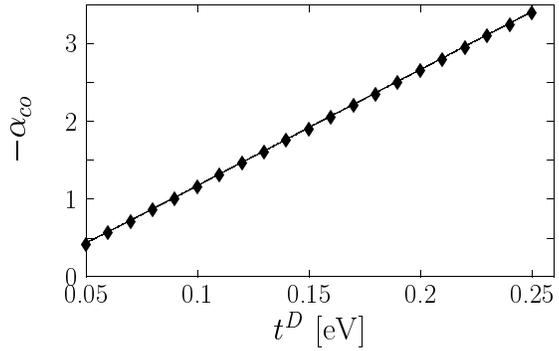,width=8cm}}
\caption{The absolute value of the oxygen isotope exponent ($-\alpha_{co}$)
  as a function of the bare DE hopping parameter $t^D$. 
  Calculations are performed for
  the parameters, $\sum_{\bf q} |u_{\bf q}|^2 = 1.2$, $V = 0.018$eV, 
  $T_{CO} = 150$K, $\omega = 0.07$eV.}
\label{alpha}
\end{figure}

The isotope induced crossover from a metallic to a CO insulating ground state
\cite{Zhao3,Ibarra,Babu} can be explained 
in the same way. The heavier ion mass of $^{18}$O reduces $\tilde{t}$, 
and so both $T_C$ and 
the insulator to metal transition temperature $ T_{MI}$ decrease.
In contrast, $ T_{CO}$ increases with decreasing $\tilde{t}$,
so that the CO insulating ground state becomes stabilized 
by the isotope exchange.
One can also understand the anomalous isotope effects observed 
in the CO phases of Nd$_{0.5}$Sr$_{0.5}$MnO$_3$ and
La$_{0.5}$Ca$_{0.5}$MnO$_3$ under the strong magnetic field \cite{Zhao4,Zhao5}.
In the presence of  the magnetic field, 
$\tilde{t}^D$ increases due to the DE term 
$\langle \cos\frac{\theta}{2}\rangle$ and this will in turn reduce $ T_{CO}$.
Under this circumstance, the isotope substitution of $^{18}$O
reduces $\tilde{t}^D$ much more rapidly than the case without the
magnetic field because of larger DE term, 
and so enhances the isotope exponent $\alpha_{co}$.
This effect is clearly demonstrated in  Fig.~\ref{alpha} which plots 
$-\alpha_{co}$ as a function of the bare DE hopping parameter $t^D$. 
The magnitude of $\alpha_{co}$ becomes larger with increasing $t^D$,
which corresponds to the case of applying the magnetic field.
Direct comparison with the experiments may be possible, once the
precise functional relation between $t^D$ and the 
magnetic field is known.
The variation of $-\alpha_{co}$ with $t^D$ and the
corresponding physical parameters seem 
qualitatively consistent with observations \cite{Zhao4,Zhao5},
indicating that the present argument describes the main physics
of isotope effects in the CO phase of RAMO.

It is noteworthy that similar arguments were provided 
recently by Babushkina {\it et al.} \cite{Babu2}
in analyzing their isotope induced metal-insulator transition
in (La$_{0.175}$Pr$_{0.525}$)Ca$_{0.3}$MnO$_3$.
In their model, the isotope effect is caused by the 
modification of the effective hopping integrals due to 
the change of interatomic distance by lattice vibrations.
They obtained the relation between
$T_{CO}$ and the effective hopping parameter, and speculated
that the large isotope effect arises from the hopping parameter
that is very close to a certain critical value, $t_c=zV/2$.
In fact, this result corresponds to the
present one in the narrow band limit $\frac B{zV} \ll 1$.
So their model is valid only in this limit. Further, they did not consider 
the polaron effect, so that the isotope mass dependence of the effective 
hopping integral is weak, $\sim \frac{1}{\sqrt{M}}$, as opposed to the 
present case $\sim \exp\left[-\sqrt{M}\right]$.
Accordingly, their model is difficult to apply to the 
analysis of the anomalous isotope effects in the presence of the 
magnetic field \cite{Zhao4,Zhao5}.

To summarize, we have studied the isotope effects both in
the metallic phase and  the CO phase of RAMO, using the combined 
model of the double exchange and the interacting lattice polaron mechanism. 
We have investigated the charge ordering transition in 
the mean-field approximation, and obtained the relation between
$T_{CO}$ and the effective hopping parameter. 
From this, we have shown that the various isotope effects including the
magnetic field induced enhancement of $\alpha_{co}$ 
in the CO phase of half-doped manganites can be explained in a natural way.

Acknowledgements$-$
Helpful discussions with J.D. Lee are greatly appreciated.
This work was supported by the KOSEF (96-0702-0101-3),
and in part  by the BSRI program of the KME (BSRI-98-2438).

\newpage

\begin{references}
\bibitem{Skr} Detailed calculations will be presented elsewhere.
\end{references}
\end{document}